# Simulating Parallel Algorithms in the MapReduce Framework with Applications to Parallel Computational Geometry

Michael T. Goodrich
Department of Computer Science
University of California, Irvine

**Abstract**

In this paper, we describe efficient MapReduce simulations of parallel algorithms specified in the BSP and PRAM models. We also provide some applications of these simulation results to problems in parallel computational geometry for the MapReduce framework, which result in efficient MapReduce algorithms for sorting, 1-dimensional all nearest-neighbors, 2-dimensional convex hulls, 3-dimensional convex hulls, and fixed-dimensional linear programming. For the case when reducers can have a buffer size of $B = O(n^\epsilon)$, for a small constant $\epsilon > 0$, all of our MapReduce algorithms for these applications run in a constant number of rounds and have a linear-sized message complexity, with high probability, while guaranteeing with high probability that all reducer lists are of size $O(B)$.

## 1 Introduction

The ***MapReduce framework*** [12, 13] is a programming paradigm for designing parallel and distributed algorithms, which can be easily implemented in cloud computing environments and server clusters (e.g., see [24]). It provides a simple programming interface that is specifically designed to make it easy for a programmer to design an efficient parallel program that can efficiently perform a data-intensive computation. This framework is gaining wide-spread interest in both theory and systems domains, in that it is motivating new models for parallel computing [15, 21] while also being used in Google data centers and as a part of the open-source Hadoop system [29] for server clusters.

In this framework, a computation is specified as a sequence of map, shuffle, and reduce steps that operate on a set $X = \{x_1, x_2, \ldots, x_n\}$ of values:

- A ***map step*** applies a function, $\mu$, to each value, $x_i$, to produce a key-value pair, $(k_i, v_i)$. To allow for parallel execution, the computation of the function, $\mu(x_i) \to (k_i, v_i)$, must depend only on $x_i$.
- A ***shuffle step*** collects all the key-value pairs produced in the previous map step, and produces a set of lists, $L_k = (k; v_{i_1}, v_{i_2}, \ldots)$, where each such list consists of all the values, $v_{i_j}$, such that $k_{i_j} = k$ for a key $k$ assigned in the map step.
- A ***reduce step*** applies a function, $\rho$, to each list, $L_k = (k; v_{i_1}, v_{i_2}, \ldots)$, formed in the shuffle step, to produce a set of values, $y_{j_1}, y_{j_2}, \ldots$. The reduction function, $\rho$, is allowed to be defined sequentially on $L_k$, but should be independent of other lists $L_{k'}$ where $k' \neq k$.

Outputs from a reduce step can, in general, be used as inputs to another round of map-shuffle-reduce steps. So we allow the output values from a reduce step can be either ***final values***, which are included in the final output of the algorithm, or they can be *intermediate values*, which are used as input for another round of map-shuffle-reduce steps. Typically, only values output in the last round of the algorithm are labeled as final. Thus, a typical MapReduce computation is described as a sequence of map-shuffle-reduce steps that perform a desired action and produce the output after the last reduce step.

For example, consider an often-cited MapReduce algorithm to count all the instances of words in a document. Given a document, $D$, we define the set of input values $X$ to be all the words in the document and we then proceed as follows:

1. Map: For each word, $w$, in the document, map $w$ to $(w, 1)$.
2. Shuffle: collect all the $(w, 1)$ pairs for each word, producing a list $(w; 1, 1, \ldots, 1)$, noting that the number of 1's in each such list is equal to the number of times $w$ appears in the document.
3. Reduce: scan each list $(w; 1, 1, \ldots, 1)$, summing up the number of 1's in each such list, and output a pair $(w, n_w)$ as a final output value, where $n_w$ is the number of 1's in the list for $w$.

This single-round computation clearly computes the number of times each word appears in $D$.

## 1.1 Evaluating MapReduce Algorithms

There are several metrics that one can use to measure the efficiency of a MapReduce algorithm over the course of its execution, including the following:

- $t$: the **number of rounds** of map-shuffle-reduce that the algorithm uses.
- $n_{i,1}, n_{i,2}, \ldots$: the **reducer I/O sizes** for round $i$, so that $n_{i,j}$ is the size of the inputs and outputs for reducer $j$ in round $i$.
- $M_i$: the **message complexity of round** $i$ of the algorithm, that is, the total size of the inputs and outputs for reducers in round $i$, that is, $M_i = \sum_j n_{i,j}$. We can also define a **message complexity**, $M = \sum_{i=1}^{t} M_i$, for the entire algorithm.
- $r_i$: the **internal running time** for round $i$, which is the maximum internal running time taken by a reducer in round $i$, where we assume $r_i \geq \max_j \{n_{i,j}\}$, since a reducer must have a running time that is at least the size of its inputs and outputs. We can also define an **internal running time**, $r = \sum_{i=1}^{t} r_i$, for the entire algorithm, as well.
- $B$: the **buffer size** for reducers, that is, the maximum size of the working memory needed by a reducer to process its inputs and outputs (in addition to the storage used for the input itself), taken across all $t$ rounds of the algorithm.

We can make a crude calibration of a MapReduce implementation using the following additional parameters:

- $L$: the latency $L$ of the shuffle network, which is the number of steps that a mapper or reducer has to wait until it receives its first input in a given round.
- $b$: the bandwidth of the shuffle network, which is the number of elements in a MapReduce computation that can be delivered by the shuffle network in any time unit.

Given these characterizations, then, the total running time, $T$, of an implementation of a MapReduce algorithm can be crudely characterized as follows:

$$\begin{aligned} T &\in O\left(\sum_{i=1}^{t}(r_i + L + M_i/b)\right) \\ &= O(r + tL + M/b), \end{aligned}$$

which we call the ***MapReduce running time***. For example, given a document $D$ of $n$ words, the simple word-counting MapReduce algorithm given above has a worst-case performance of $t$ being 1, $M$ being $O(n)$, and $r$ being $O(n)$; hence, its worst-case time performance is $O(n)$. Unfortunately, such performance could be quite common, as the frequency of words in a natural-language document tend to follow Zipf's law, so that some words appear quite frequently, and the running time of the simple word-counting algorithm is proportional to the number of occurrences of the most-frequent word. For instance, in the Brown Corpus [23], the word "the" accounts for 7% of all word occurrences[1].

Note, therefore, that focusing exclusively on the number of rounds in a MapReduce algorithm can actually lead to inefficient algorithms. For example, if we focus only on the number of rounds, $t$, then the most efficient algorithm would always be one that maps all the inputs to a single key and then has the reducer for this key perform a standard sequential algorithm to solve the problem. This approach would run in one round, but it would not use any parallelism. Specifically, it would have a running time of $O(\tau(n)+L+n/b)$, where $\tau(n)$ is the running time of the sequential algorithm; hence, this MapReduce algorithm would be only as efficient as the best sequential algorithm.

## 1.2 Memory-Bound MapReduce Algorithms

Possibly as a way to steer algorithm designers away from such misplaced optimizations, some recent formalizations of the MapReduce paradigm have focused primarily on optimizing the round complexity bound, $t$, while restricting $B$, the buffer size for reducers. Karloff *et al.* [21] define their MapReduce model, MRC, so that $B$ is restricted to be $O(n^{1-\epsilon})$ for some small constant $\epsilon > 0$, and Feldman *et al.* [15] define their model, MUD, so that $B$ is restricted to be $O(\log^d n)$, for some constant $d \geq 0$, and reducers are further required to process their inputs in a single pass. These restrictions limit the feasibility of the trivial sequential algorithm for solving a problem in the MapReduce framework and instead compel algorithm designers to setup computations so no reducer is given an input that is larger than $B$ in any given round.

In this paper, we take a somewhat more general viewpoint, in that we allow for arbitrary MapReduce computations, but we evaluate the efficiency of these algorithms in terms of their MapReduce running times with respect to the reducer I/O sizes. In addition, we make some simplifying assumptions that allow the characterization of a MapReduce computation to be more concise. In particular, suppose we can define the reducer buffer size, $B$, in such a way that we can restrict our algorithm so that $n_{i,j}$ is $O(B)$, for each round $i$ and reducer $j$, which we refer to as a computation being a ***memory-bound MapReduce algorithm***. Thus, if the running time of a reducer is dominated by the time it takes to route its inputs and outputs (which should be common), then the running time of a memory-bound MapReduce algorithm can be characterized as being

$$O(t(B+L) + M/b).$$

In other words, independent of the values of $L$ and $b$, we can capture the essential efficiency of a typical memory-bound MapReduce algorithm by characterizing its round complexity and message complexity[2]. Of course, in a memory-bound MapReduce algorithm, such bounds for $t$ and $M$ may depend on $B$, but that is fine, for similar characterizations are common in the literature on external-memory algorithms (e.g., see [2, 4, 5, 27, 28]).

---

[1] http://en.wikipedia.org/wiki/Zipf's_law

[2] These measures correspond naturally with the ***time*** and ***work*** bounds used to characterize PRAM algorithms (e.g., see [20]).

## 1.3 Our Results

We provide a simulation result that shows that any Bulk-Synchronous Parallel (BSP) algorithm [25] running in $T$ steps with a memory of size $N$ and $P \leq N$ processors can be simulated with a memory-bound MapReduce algorithm in $T$ rounds and message complexity $O(TN)$ with reducer buffers of size $B \in O(N/P)$, with high probability. In addition, to facilitate such a simulation, we provide a solution to the problem of indexing an input set for MapReduce, which involves identifying each input $x_i$ with the index $i$, which is not, in general, included in the input format for a MapReduce algorithm. Our indexing method involves a randomized $B$-**ary distribute-and-combine** technique.

We also give simulation results showing that any CRCW PRAM algorithm running in $T$ steps with $P$ processors on a memory of size $N$ can be simulated with a memory-bound randomized MapReduce algorithm in $O(T \log_B P)$ rounds with $O(T(N + P) \log_B N)$ message complexity, with high probability. This latter simulation result holds for any version of the CRCW PRAM model, including the $f$-CRCW PRAM, which involves the computation of a commutative semigroup operator $f$ on concurrent writes to the same memory location, such as in the Sum-CRCW PRAM [14]. Our approach in this case uses a technique we call the **invisible B-tree** method, as it can be viewed as placing a $B$-tree, $\mathcal{T}_j$, "on top" of each PRAM memory cell $j$ and routing reads and writes to $j$ via $\mathcal{T}_j$, without ever explicitly constructing any B-tree.

In addition to providing these simulation results, we show how to apply our approach to solve several parallel computational geometry problems, including sorting, 1-dimensional all-nearest neighbors, the convex hull problem in 2- and 3-dimensions, and fixed-dimensional linear programming. In particular, we give memory-bound MapReduce algorithms for these problems that run in $O(\log_B N)$ rounds with message complexities that are $O(N \log_B N)$, with high probability, for inputs of size $N$. Thus, for the case when $B$ is $\Theta(N^\epsilon)$, for a small constant $\epsilon > 0$, these parallel computational geometry algorithms run in a constant number of rounds with linear message complexity in the memory-bound MapReduce model. Note that, in most applications, we would indeed expect the buffer size, $B$, for reducers to be $O(N^\epsilon)$, since the MapReduce framework is designed to solve problems that are too large to fit in the memory space of a single computer but not so large as to be exponentially larger than the memory space of a single computer. In such cases, the round complexities and message complexities of our computational geometry algorithms are within constant factors of optimal.

## 2 Indexing the Input

The input to a BSP or PRAM algorithm is implicitly indexed, so that, for each input $x_i$, we know the value of $i$. The input to a MapReduce computation, on the other hand, is typically not indexed. So a necessary first step to being able to simulate a BSP or PRAM algorithm in the MapReduce framework is to index the input. We describe a method for performing this indexing in this section.

Our method does not require that all mappers know the value of $N$, the size of the input, but we do assume that there is a value, $\hat{N}$, such that $\hat{N}$ is known to all mappers and $N \leq \hat{N} \leq bN^c$, for some constants $b \geq 1$ and $c \geq 1$. Our method is as follows.

Imagine that we have a B-tree, $\mathcal{T}$, of height $L = \lceil 3 \log_B \hat{N} \rceil$, so that there are $\hat{N}^3$ leaves in $\mathcal{T}$. Furthermore, let us label the nodes in $\mathcal{T}$ so that the $i$th node on level $l$ is labeled $[l, i]$. Thus, the parent of a non-root node labeled $[l, i]$ is $[l-1, \lfloor i/B \rfloor]$. In other words, given a node $v$ in $\mathcal{T}$, we can identify the parent, $p(v)$, of $v$ in $\mathcal{T}$ based solely on the label for $v$.

**Initialization step.** For each input value $x_i$, map $x_i$ to the tuple $(v_i, x_i, 1)$, where $v_i$ is a randomly-chosen leaf in $\mathcal{T}$. That is, we choose for $x_i$ a node $v_i$ that is labeled $[L, r_i]$, where $r_i$ is a random number chosen uniformly in the range $[1, \hat{N}^3]$. Thus, the probability that we choose for $x_i$ a node $v_i$ that is chosen by any other $x_j$ is at most $1/N^2$; hence, the probability that any collisions occur is at most $1/N$. Moreover, if we want to make this probability even smaller, we can choose $r_i$ in a larger range that is polynomial in $\hat{N}$. So let us assume there are no collisions (and we can, in fact, even allow up to $B$ collisions with a small modification to the algorithm, and the event that there are more than $B$ collisions is very small). We then perform a bottom-up phase followed by a top-down phase.

**Bottom-up phase.** Repeat the following steps for $L$ times:
1. Map: perform the identity map $(v, x, s)$ to $(v, x, s)$ for each item.
2. Reduce: input is a list $(v; (x_1, s_1), (x_2, s_2), \ldots, (x_k, s_k))$. Output $(p(v), v, \sum_{i=1}^{k} s_i)$. In addition, output $(v, x_i, s_i)$ for each $(x_i, s_i)$ in this list.

**Top-down phase.** Repeat the following steps for $L$ times:
1. Map: perform the identity map $(v, x, s)$ to $(v, x, s)$ for each item.
2. Reduce: input is a list $(v; (x_1, s_1), (x_2, s_2), \ldots, (x_k, s_k))$. If $v$ is the the root of $\mathcal{T}$, then set $s = 0$. Otherwise, find the pair $(x_i, s_i)$ such that $v = x_i$, and set $s = s_i$, and remove it from the list for $v$. Inductively, the value $s$ is the sum of all items "to the left" of $v$. If $v$ is a leaf, then, for the one item, $(x_1, s_1)$, output $(v, x_i, s + s_1)$ as a final value. Otherwise, if $v$ is not a leaf, then output $(x_i, x_i, s + \sum_{j=1}^{i-1} s_j)$ as an intermediate value (to be used as input to the map step in the next iteration). In this latter case, the sum being passed down satisfies the induction hypothesis.

This algorithm computes the prefix sums of all the initial values associated with the inputs, $x_i$ in $X$, which if they are all 1's, computes the index $i$ of each $x_i$. Thus, we have the following.

**Theorem 2.1:** *Given a set $X$ of $N$ values, we can compute a random indexing of the values in $X$ from 1 to $N$, or a random ordering of the items in $X$ and a set of prefix sums based on this ordering, in the memory-bound MapReduce framework with a number of rounds that is $O(\log_B N)$ and a message complexity that is $O(N \log_B N)$ with high probability.*

## 3 Simulating BSP Algorithms and MapReduce Sorting

Karloff *et al.* [21] provide an optimal simulation of any EREW PRAM algorithm in the MapReduce framework using reducers with constant-size memories. In particular, they show how to simulate an EREW PRAM algorithm that runs in $T$ steps using $P$ processors and $N$ memory cells with a memory-bound MapReduce algorithm that runs in $T$ rounds with message complexity $O(T(N+P))$, with constant-size reducer buffers. While this is useful, we show in this section that we can often achieve even more efficient MapReduce algorithms by simulating a BSP algorithm and taking better advantage of the full breadth of the memory of size $B$ that reducers can use to process their inputs. For example, using the Karloff *et al.* result to simulate the best EREW PRAM convex hull algorithm [10] results in a MapReduce algorithm that runs in $O(\log N)$ time with message complexity $O(N \log N)$ to compute the convex hull of a set of $N$ points in the plane. Our BSP-based approach, on the other hand, results in a MapReduce convex hull algorithm that runs in $O(\log_B N)$ time with message complexity $O(N \log_B N)$, with high probability.

In the BSP model [25], we are given an input of size $n$ allocated to $p$ processors so that each processor is assigned at most $m = \lceil n/p \rceil$ input items and a computation is specified as a series of super-steps, each of which involves each processor performing an internal computation and then sending a set of up to $m$ messages to other processors in batch. This model is similar to other parallel models, incidentally, including the XPRAM [26], EREW phase-PRAM [17], and bulk-synchronous LogP model [11, 22].

Consider now a BSP algorithm, $\mathcal{A}$, specified on an input that is initially stored in a memory of size $n$ assigned evenly to $p$ processors. If the input to our algorithm is not already indexed, then we index it using the method of Theorem 2.1. This allows us to assume that we can define an initial state of the BSP algorithm by an indexed set of processors items, $(1, P_1), (2, P_2) \ldots, (p, P_p)$ and an indexed set of initialized memory cells $(1, M_{1,1}), (1, M_{1,2}), \ldots, (p, M_{p,m})$, such that $M_{i,j}$ is the $j$th memory cell assigned to processor $i$. Note: we assume each $M_{i,j}$ is a single-word memory cell (together with an internal address field, $j$) and $P_i$ is a constant-size record that specifies the internal state of processor $i$, including its program counter and internal registers. These are the items that we consider as the inputs to our MapReduce simulation algorithm.

We assume inductively that there may be a batched set of messages that are to be included in the input to the $k$th super-step of the algorithm $\mathcal{A}$. These messages are specified as a tuple $(i, C)$, where $i$ is the destination processor and $C$ is the contents of the message. Each round of our simulation corresponds to a super-step in the BSP algorithm $\mathcal{A}$ and is as follows.

- Map: Perform the identity map for each indexed memory cell, $(i, M_{i,j})$, each indexed processor state, $(i, P_i)$, and each message, $(i, m)$.
- Reduce: input is $(i; P_i, M_{i,1}, M_{i,2}, \ldots, M_{i,m}, C_1, \ldots, C_l)$. Perform the computation for processor $i$, as specified by its state, $P_i$, for this super-step of $\mathcal{A}$, and the state of its memory cells, $M_{i,1}, \ldots, M_{i,m}$ and the messages it receives, $C_1, \ldots, C_l$, in this super-step of $\mathcal{A}$. This activity generates a new state pair, $(i, P_i')$, indicating the new state for processor $i$, as well as new indexed values for its internal memory cells, $(i, M_{i,1}'), (i, M_{i,2}'), \ldots, (i, M_{i,m}')$. In addition, the computation for $i$ in this super-step defines $m' \leq m$ messages, $(j_1, C_1), \ldots, (j_{m'}, C_{m'})$. These are the outputs from the reducer for the list $i$. Note that the entire reducer computation uses $O(m)$ internal memory and has inputs and outputs of size $O(m)$.

Performing the MapReduce simulation for $T$ steps of a BSP algorithm $\mathcal{A}$ results in the final state of $\mathcal{A}$, which we then identify as the output. Thus, we have the following.

**Theorem 3.1:** *Given a BSP algorithm $\mathcal{A}$ that runs in $T$ super-steps with a total memory size $N$ using $P \leq N$ processors, we can simulate $\mathcal{A}$ using $T$ rounds and message complexity $O(TN)$ in the memory-bound MapReduce framework with reducer buffer size bounded by $B = \lceil N/p \rceil$.*

## 4  MapReduce Sorting and Convex Hulls

We list some immediate implications of Theorem 3.1 for MapReduce sorting and convex hulls in this section.

**MapReduce Sorting.** In the MapReduce version of the classic sorting problem, we are given a set $X$ of comparable items and asked to compute for each $x$ in $X$ the number of items in $X$ that are less than or equal to $x$. Since we can compute a random indexing on $X$, by Theorem 2.1, we can assume without loss of generality that the elements in $X$ are distinct. Thus, we can also use sorting to solve the 1-dimensional ***all-nearest neighbors*** problem, in which, for each $x$ in $X$, we are asked to find the smallest element in $X$ that is larger than $x$. We have the following.

**Corollary 4.1:** *Given a set $X$ of $N$ indexed comparable items, we can sort $X$ or solve the 1-dimensional all nearest neighbors problem for $X$ with $O(\log_B N)$ rounds and $O(N \log_B N)$ message complexity in the memory-bound MapReduce model. Moreover, this performance is optimal up to constant factors for this model.*

**Proof:** Goodrich [19] provides a BSP sorting algorithm that runs in $O(\log_m n)$ time for sorting $n$ items on $p$ processors using $O(n)$ memory cells, where $m = \lceil n/p \rceil$. Defining $p = \lceil N/B \rceil$ and simulating this algorithm as determined by Theorem 3.1 gives us the upper bounds. Goodrich also shows that his algorithm is asymptotically optimal for sorting in the BSP model, and this lower-bound proof carries over to the memory-bound MapReduce model. ∎

**MapReduce Convex Hulls.** In the MapReduce planar convex hull problem, we are given a set $S$ of $n$ points in the plane, and we want to compute a representation of the convex hull of $S$. In this case, such a representation could be an indexed listing of the vertices of the convex hull in counter-clockwise order.

**Corollary 4.2:** *Given a set $S$ of $N$ points in the plane, we can compute a representation of the convex hull of $S$ in the memory-bound MapReduce model in $O(\log_B N)$ rounds and with $O(N \log_B N)$ message complexity, with high probability.*

**Proof:** Goodrich [18] provides a BSP convex hull algorithm that runs in $O(\log_m n)$ time for computing the two-dimensional convex hull of $n$ points in the plane on $p$ processors using $O(n)$ memory cells, where $m = \lceil n/p \rceil$, with high probability. Defining $p = \lceil N/B \rceil$, applying Theorem 2.1 to index the input points, and simulating this algorithm as determined by Theorem 3.1 gives us the claimed result. ∎

For the three-dimensional convex hull problem, the input is a set $S$ of points in $\mathbf{R}^3$ for which we want to compute a representation of the convex hull. Such a representation could be an identification and label of each triangular face on the convex hull, together with the labels of its three adjacent triangular faces.

**Corollary 4.3:** *Given a set $S$ of $N$ points in $\mathbf{R}^3$, we can compute a representation of the convex hull of $S$ in the memory-bound MapReduce model in $O(\log_B N)$ rounds and with $O(N \log_B N)$ message complexity, with high probability.*

**Proof:** Goodrich [18] provides a BSP convex hull algorithm that runs in $O(\log_m n)$ time for computing the 3-dimensional convex hull of $n$ points on $p$ processors using $O(n)$ memory cells, where $m = \lceil n/p \rceil$, with high probability. Defining $p = \lceil N/B \rceil$, applying Theorem 2.1 to index the input points, and simulating this algorithm as determined by Theorem 3.1 gives us the claimed result. ∎

## 5  Simulating CRCW PRAM Algorithms via Invisible B-trees

Let us now consider how we could simulate a CRCW PRAM algorithm, $\mathcal{A}$, where we allow for the most powerful version of this model in that we allow concurrent writes to be resolved through the computation of a commutative semigroup operation $f$ on all colliding write values, such as Sum, Min, Max, etc.

By Theorem 2.1, we can assume that the input to our simulation is specified by an indexed set of $p$ processor items, $(1, P_1), (2, P_2) \ldots, (p, P_p)$, and an indexed set of initialized memory cells, $(1, M_1), (1, M_2), \ldots, (N, M_N)$, where $N$ is the total memory size used by $\mathcal{A}$. Note: we assume each $M_i$ is a single-word memory cell (together with an internal address field, $i$) and $P_i$ is a constant-size record that specifies the internal state of processor $i$, including its program counter and internal registers.

The main challenge in simulating the algorithm $\mathcal{A}$ in the memory-bound MapReduce model is that there may be as many as $p$ reads and writes to the same memory cell in any given step and $p$ can be significantly larger than $B$, the buffer size for reducers. Thus, we need to have a way to "fan in" these reads and writes. We do this using an ***invisible B-tree*** technique, where we imagine that there is a different invisible B-tree rooted at each memory cell that has the set of processors as its leaves. Intuitively, our simulation algorithm involves routing reads and writes up and down these $N$ invisible B-trees. We view them as "invisible," because we don't actually maintain them explicitly, since that would require $\Theta(pN)$ additional memory cells.

As described above in Section 2, any time we have a B-tree, $\mathcal{T}$, of height $L$, we can label the nodes in $\mathcal{T}$ so that the $i$th node on level $l$ is labeled $[l, i]$. Thus, there is a simple label-based way of identifying the parent, $p(v)$, of any node $v$ in $\mathcal{T}$ (based solely on the label for $v$) and similarly a simple label-based way of identifying all the children of $v$ as well. In our CRCW PRAM simulation, however, we have $p$ B-trees, so we index each node as a pair $(j, v)$, where $v$ is a node label (as above) in the $j$th invisible B-tree.

We view the computation specified in a single step in the algorithm $\mathcal{A}$ as being composed of a read sub-step, followed by a constant-time internal computation, followed by a write sub-step. We simulate each such step as follows, for $L = \lceil \log_B p \rceil$.

**Bottom-up read phase.** At the beginning of the read phase, we assume that we have a set of up to $p$ read requests of the form $(j, i)$, which indicates that processor $i$ would like to read the contents of memory cell $j$. We initially map the pair $(j, i)$ to the pair $((j, v), i)$, where $v$ is the $i$th leaf node in an $L$-depth B-tree, so $(j, v)$ is the $i$th leaf node in the invisible B-tree for memory cell $j$. We perform the following sequence of map-reduce steps for $L$ times.

1. Map: perform the identity map $((j, v), i)$ to $((j, v), i)$ for each read request. We also perform identity maps for each $(i, P_i)$ and $(i, M_i)$.

2. Reduce1: input is a list $((j, v); i_1, \ldots, i_k)$. If $v$ is not the root of the $j$-th B-tree, then output $((j, p(v)), (j, v))$, to indicate that the node $(j, v)$ would like to read the contents of the $j$th memory cell. If $v$ is the root, then we output $(j, v)$. In either case, we also output $((j, v), i_l)$ for each $i_l$ in this reduce list.

3. Reduce2: input is a list of the form $(j; P_j, M_j)$ or $(j; P_j, M_j, v)$. Output $(j, P_j)$, $(j, M_j)$, and, if there is a $v$, then output $((j, v), M_j)$ to indicate that we are ready to send the contents of $M_j$ to the $j$th invisible B-tree. Note that this latter case occurs only at the end of this bottom-up read phase.

**Top-down read phase.** Repeat the following steps for $L$ times:

1. Map: perform the identity map for each $((j, v), x)$. We also perform identity maps for each $(i, P_i)$ and $(i, M_i)$.

2. Reduce1: input is a list $((j, v); M_j, x_1, \ldots, x_k)$. If $v$ is not a leaf, then, for each $x_i$ in this list, output $((j, x_i), M_j)$. If $v$ is a leaf, then output $(x_i, M_j)$, for each $x_i$ in this list.

3. Reduce2: input is a list of the form $(j; P_j, M_j)$ or $(j; P_j, M_j, M_k)$. Output $(j, M_j)$, and, if there is no $M_k$, output $(j, P_j)$. If there is a $M_k$, then perform the computation for $P_j$ in response to its requested memory memory cell, $M_k$. This results in a new state for processor $j$, which we output as $(j, P'_j)$. It may also involve a write-request where processor $j$ wants to write some new contents to some memory cell $l$, which we output as $((l, v), M_l)$, where $(l, v)$ is the leaf in the $l$th B-tree (for memory cell $M_l$). Note that this latter case occurs only at the end of this top-up read phase.

After the top-down read phase completes, then we switch over to the bottom-up write phase.

**Bottom-up write phase.** We perform the following sequence of map-reduce steps for $L$ times.

1. Map: perform the identity map $((j, v), x)$ to $((j, v), x)$ for each write request. We also perform identity maps for each $(i, P_i)$ and $(i, M_i)$.

2. Reduce1: input is a list $((j, v); M_1, \ldots, M_k)$, which indicates that processors would like to write the values $M_1$ to $M_k$ to memory cell $j$. So we compute $M'$, which is the result of applying the combining function, $f$, to the values $M_1$ to $M_k$. If $v$ is not the root of the $j$-th B-tree, then we output $((j, p(v)), M')$. If $v$ is the root, then we output $(j, (M', \text{new}))$.

3. Reduce2: input is a list of the form $(j; P_j, M_j)$ or $(j; P_j, M_j, (M', \text{new}))$. We output $(j, P_j)$, and, if there is no $(M', \text{new})$ pair, then we output $(j, M_j)$. Otherwise, if there is an $(M', \text{new})$, then we output $(j, M')$. Note that this latter case occurs only at the end of this bottom-up write phase.

When we have completed this write phase, then we are inductively ready for simulating the next step in the PRAM algorithm. Thus, we have the following.

**Theorem 5.1:** *Given an algorithm $\mathcal{A}$ in the CRCW PRAM model, with write conflicts resolved according to a commutative semigroup function, $f$, such that $\mathcal{A}$ runs in $T$ steps using $P$ processors and $N$ memory cells, we can simulate $\mathcal{A}$ in the memory-bound MapReduce framework in $O(T \log_B P)$ rounds and with $O(T(N + P) \log_B P)$ message complexity.*

This implies the following.

**Corollary 5.2:** *Given a set $S$ of $N$ linear constraints in $\mathbf{R}^d$, for a fixed constant $d \geq 1$, and a linear optimization function, we can solve this linear programming problem in the memory-bound MapReduce model in $O(\log_B N)$ rounds and with $O(N \log_B N)$ message complexity, with high probability.*

**Proof:** Alon and Megiddo [3] provide a CRCW PRAM linear programming algorithm that runs in $O(1)$ time for fixed dimensions, using $N$ processors and $O(N)$ memory cells, with very high probability. Applying Theorem 2.1 to index the input points, and simulating this algorithm as determined by Theorem 5.1 gives us the claimed result. ∎

## 6 Conclusion

We have given efficient simulations of BSP and PRAM parallel algorithms in the memory-bound MapReduce framework. Moreover, our simulations lead to efficient MapReduce algorithms for sorting, 1-dimensional all nearest-neighbors, 2-dimensional convex hulls, 3-dimensional convex hulls, and fixed-dimensional linear programming. Admittedly, the algorithmic path we take in this paper, of solving a problem in the MapReduce framework by simulating a BSP or PRAM algorithm, can result in asymptotically-efficient methods that may not be the simplest of algorithms, and such is the case for the algorithms we have given here. Thus, a natural set of open problems is to design simple efficient algorithms directly in the memory-bound MapReduce model for the problems we study in this paper.

In addition, the memory-bound MapReduce algorithms we discuss in this paper make explicit use of knowledge of the reducer buffer size, $B$. It would be interesting to see if one could find a way to express efficient memory-bound MapReduce algorithms that are oblivious to this value, much in the same way that cache-oblivious algorithms are oblivious to the size of blocks in an external-memory context (e.g., see [1, 8, 6, 7, 9, 16]).

## Acknowledgments

We would like to thank Tyson Condie and Siddharth Suri for several helpful email exchanges regarding the MapReduce framework. This research was supported in part by the National Science Foundation under grants 0724806, 0713046, and 0847968, and by the Office of Naval Research under MURI grant N00014-08-1-1015.